\documentstyle[11pt,aaspp4]{article}
\slugcomment{accepted by ApJ Jan 06 2000}
\lefthead{Leggett et al.}
\righthead{Spectral Energy Distributions for M--Dwarfs}
 
\begin{document}
 
\title{Spectral Energy Distributions for Disk and Halo M--Dwarfs}

\author{S.K. Leggett}
\affil{Joint Astronomy Centre, University Park, Hilo, HI 96720
\nl skl@jach.hawaii.edu}
 
\author{F. Allard} 
\affil{CRAL, Ecole Normale Superieure, 46 Allee d'Italie, Lyon, 69364 France
\nl  fallard@cral.ens-lyon.fr } 

\author{Conard Dahn}
\affil{Naval Observatory Flagstaff Station, Flagstaff AZ 86002-1149
\nl dahn@nofs.navy.mil}
 
\author{P.H. Hauschildt}
\affil{Department of Physics and Astronomy \& Center for Simulational Physics, 
\nl University of Georgia, Athens, GA 30602-2451
\nl yeti@hal.physast.uga.edu}

\author{T.H. Kerr}
\affil{Joint Astronomy Centre, University Park, Hilo, HI 96720
\nl tkerr@jach.hawaii.edu}

\and

\author{J. Rayner}
\affil{IRTF, 2680 Woodlawn Drive, Honolulu HI 96822
\nl rayner@irtf.ifa.hawaii.edu}

\begin{abstract}
We have obtained infrared (1---2.5$\mu$m) spectroscopy for 
42 halo and disk dwarfs with spectral type M1 to M6.5.
These data are compared to synthetic spectra generated by
the latest model atmospheres of Allard \& Hauschildt.
Photospheric parameters metallicity, effective temperature
and radius are determined for the sample.

We find good agreement between observation and theory except for
known problems due to incomplete
molecular data for metal hydrides and H$_2$O.
The metal--poor M subdwarfs are well matched by the models 
as oxide opacity sources are less important in this case.
The derived effective temperatures for the sample range from 
3600~K to 2600~K; at these temperatures grain formation
and extinction are not significant in the photosphere.
The derived metallicities range from solar to one-tenth solar. 
The radii
and effective temperatures derived agree well with  recent
 models of low mass stars.  
 
The spectra are available in electronic form, on request.

\end{abstract}
%\keywords{}
%\keywords{globular clusters,peanut clusters,bosons,bozos}
 
\section{Introduction}

Until very recently the observational data for low--mass stars could not be 
well reproduced by
synthetic spectra or photometry.  The cool, high--pressure 
atmospheres are difficult to model, due in particular to complex opacity 
sources: strong molecular bands and, for the halo stars and very low--mass 
objects, pressure--induced molecular hydrogen opacity (see e.g.
\cite{bor97}). The situation is now much
improved as demonstrated by for example the ``NextGen'' models of 
\cite{all98b} (see also \cite{all97}).
The observational side of the study of low--mass stars  has also changed
remarkably with a large increase in the known number of
low--mass stars, brown dwarfs, and even giant planets.  

For the last several years we have been obtaining infrared spectra for 
a sample of halo and disk stars approaching and even below the 
stellar/sub--stellar boundary.   This
work extends the similar spectroscopic study presented by \cite{l96}
to more metal--poor and lower mass regimes, and builds on the photometric study
published recently by us (\cite{l98}).  The spectra are
compared to synthetic spectra generated from an improved version of 
Allard \& Hauschildt's NextGen model atmospheres (\cite{all98b}). 
In this paper we will present the results for the hotter stars in the sample,
those with effective temperature higher than 2500~K, where neither grain 
condensation or extinction is significant.  A subsequent paper will present our
results for the cooler objects.

In \S 2 we describe the various instruments that have been used
for this project.  The target sample is described in \S 3 where we
illustrate the likely range of metallicity and effective temperature
through color--color and magnitude--color diagrams.  \S 4 gives the
observational results in the form of selected sets of spectra as well as
integrated fluxes and bolometric magnitudes. \S 5
gives a brief description of the models and the comparison process.  
The results of the comparison
of the data to the synthetic spectra  are presented in \S 6 and our
conclusions given in \S 7.

\section{Instrumentation}

We have been obtaining 1---2.5$\mu$m spectra of low mass stars for the
last few years, where the targets have been selected to sample a broad
range of temperature and metallicity.   Here we present the results
for forty-two of the stars, those with spectral type earlier than M7,
and effective temperature hotter than 2500~K. 
The observing runs occurred in 1999 April, 1998 January,
1997 November and 1997 April using CGS4 at UKIRT on Mauna Kea, and
1994 July using KSPEC at the University of Hawaii's 88--inch
telescope on Mauna Kea.  KSPEC is a cross--dispersed spectrometer
with a 1024$\times$1024 HgCdTe detector.  CGS4 is a grating spectrometer with
a 256$\times$256 InSb detector. Fainter stars
could be observed with CGS4 than with KSPEC, however KSPEC provided spectra
from 0.9$\mu$m to 2.4$\mu$m in a single exposure, compared to the four
grating settings required by CGS4. 

For 9 of the stars  we have obtained
optical spectra using spectrometers in Hawaii in 1996 April, and in Flagstaff 
Arizona in 1989 December and 1989 September.
In Arizona we used the B \& C spectrograph on the  72--inch Lowell
telescope and the spectrograph on the Naval Observatory Flagstaff 
Station's 40--inch telescope.   In Hawaii  we used the HARIS spectrograph at the
UH 88--inch telescope.
 
Table 1 lists the observing dates, grating and slit information, and 
resolutions, provided by these instruments.
In the cases where we did not have our own optical spectra we have
obtained spectra from the literature.  These sources 
are described below.

Data reduction was carried out in the usual way using the IRAF and Figaro
software packages.  For the infrared data, the effect of the terrestrial 
atmosphere was removed by dividing by the spectrum of a nearby early--type 
star, after removing hydrogen lines seen in this reference spectrum.  The
shape of the infrared spectrum is corrected for the known flux distribution of
the early--type 
star.  The spectral segments were individually flux calibrated using the
targets known IJHK photometry.  Each section was integrated over the 
appropriate filter profile (Cousins I or UKIRT JHK); the observed flux from 
Vega was integrated
over the same profile.  Vega was assumed to be zero magnitude at all 
wavelengths, and the target flux was scaled to match the
broadband photometry.  Where we obtained optical spectra from the literature
these were flux calibrated by us in the same way (using either an R or I filter 
as appropriate).

\section{The Sample}

\subsection{Sample Selection}

Infrared spectra were obtained for a selection of the
very low--mass stars of the halo and disk.  The halo
stars were selected
from studies of known high proper--motion stars by \cite{g97,gr97}
and \cite{m92}.  The disk stars are also known proper motion
objects, and were selected from  \cite{l92} to sample metallicity and 
a range of effective temperature. 
The targets are listed in Table 2. We give LHS or LP number 
(\cite{luy79}),
and/or Gliese or Gliese/Jahreiss number (\cite{gj91}), and/or Giclas
number (\cite{g71}), for each star.    An abbreviated RA/Dec is also 
given to aid identification. Note that LHS 421 is the 
well--known eclipsing 
binary CM Draconis (also known as  Gliese 630.1A, \cite{m66}).

The spectral types in Table 2 are taken from various sources and there may
be discrepancies or errors at the level of one sub--class.  For the halo
stars the classifications primarily are from \cite{g97} and \cite{gr97}, but 
for LHS~2045 and LHS~3390 they are based on our own
optical spectra.  For the disk stars the classifications are taken 
from \cite{g97,k95} and \cite{k97}.

The kinematic populations have been taken from \cite{l98,l92}. For a few
objects the classifications of  \cite{l92} have been updated using radial 
velocities from \cite{rhg95} and \cite{dr98}.  The classification schemes for 
young disk (YD), young/old disk (Y/O), old disk (OD), old disk/halo (O/H) 
and halo (H) are described in \cite{l92}. 

Table 2 also lists the instrumentation used for each object using the
configuration names given in Table 1.  Where we did not have our own optical 
data available we used published spectra taken
from the sources listed in the Table.  The instrumental resolutions for these
data are given in the notes to the Table.

\subsection{Photometrically Implied Properties of the Sample}

Table 3 lists distance moduli and VIJHKL$^{\prime}$ colors for the sample,
taken  primarily from the compilations by \cite{l92,l98} ---  the reader is 
asked to refer to  these papers for the data sources.  V,I for LHS 425 are 
unpublished Naval Observatory Flagstaff Station data.  New L$^{\prime}$
data have been obtained by us for LHS 1174, LHS 523 and LHS 5328 in 1999
September using the reconfigured IRCAM/TUFTI camera on UKIRT at Mauna Kea.
These data are presented in Table 3.   As shown by \cite{l98} the NextGen models
(\cite{all98b}) do a very good job of reproducing the observed photometry.
The most recent models include a new linelist for TiO (\cite{tio}) which 
has improved
the match to the energy distribution in the optical to red regime
(compare our Figure 1 to Figure 1 of \cite{l98}).
The new models also include a new linelist for H$_2$O (\cite{h2o}) however the 
calculated and observed water band strengths still show some discrepancies.
This is discussed further below.

Figures 1 to 4 show VIJHKL$^{\prime}$ color--color diagrams with model 
synthetic color sequences overlaid.  The most metal-poor stars and coolest
stars are identified.  Figure 5 shows  M$_J$:J$-$K with isochrones from
the structural models by \cite{bar97,cha97}.  These use the NextGen 
atmospheres but not the most recent versions with the improved linelists.
Nevertheless the agreement is good and we can estimate mass and metallicities
for our objects from this diagram. Note that the empirical
masses based on the \cite{hm93} scale (shown on the right axis) 
agree well with these structural models.  The
more recent mass--luminosity paper by \cite{h98} supports their previous work.

Figures 1 through 5 imply that
our sample covers a range of metallicity of about
solar to about 3\% of solar (m/H$ \sim -1.5$) and the range in effective
temperature is about 3800~K to 2400~K.  The implied mass range is 
0.3---0.1M/M$_{\odot}$ for the halo stars, 0.6---0.09M/M$_{\odot}$ for 
the disk stars.   A grid of synthetic spectra were calculated covering
this range of likely values, and more exact determinations of 
effective temperature and metallicity are presented
later as a result of  comparison to the synthetic spectra.

\section{Observational Results}

\subsection{Spectroscopic Sequences}

Figure 6 shows a representative set of spectra for approximately solar 
metallicity objects with a range of effective temperature and spectral type.
The obvious features to note, which strengthen with decreasing temperature,
are: the CO bands at 2.3$\mu$m; the water bands at around 1.4$\mu$m,
1.8$\mu$m 
and 2.4$\mu$m; the FeH band at 0.99$\mu$m; and the KI doublets at around 
1.18$\mu$m and 1.24$\mu$m.  A more complete list of spectral features seen in 
the M--dwarfs is given in \cite{l96}.

Figure 7 demonstrates the effect of metallicity at T$_{eff} \sim 3000$~K.
The shape of the infrared energy distribution for the metal--poor stars 
becomes dominated by pressure--induced H$_2$ opacity. There are no strong
absorption features seen in the subdwarfs in the infrared, but in the optical 
the hydride features
are a good indicator of a subdwarf nature --- the hydride features, especially
CaH at 0.69$\mu$m, become very strong relative to the TiO bands.

\subsection{Integrated Fluxes and Bolometric Corrections}

Table 3 gives integrated fluxes for the sample, expressed as flux at the
Earth, bolometric magnitude and intrinsic stellar luminosity.  The
integrated fluxes were
obtained by integrating our spectroscopic data over wavelength, and adding the 
flux contributions at shorter and longer wavelengths.  Some stars had gaps 
in our spectroscopic data around 1$\mu$m. The contributions from these regions
were estimated using stars  of similar spectral type
with complete spectral coverage.

The flux contributions at  wavelengths beyond 2.4$\mu$m were
calculated by deriving the flux at L$^{\prime}$ using an effective wavelength
approach, summing the contribution from the end of the K--band spectrum to
this point with a linear interpolation, and assuming a Rayleigh--Jeans tail 
beyond L$^{\prime}$. Theoretical energy distributions imply that the error in 
a Rayleigh--Jeans assumption is $\ll$1\% for this sample. For
stars without L$^{\prime}$ photometry, L$^{\prime}$ was estimated from stars of
similar J$-$K color and metallicity. 

Most of our stars have spectra available starting around 0.6$\mu$m.  For
these stars the shorter wavelength flux contribution was adopted to be  a 
simple linear extrapolation to zero flux at zero wavelength from the flux at 
0.6$\mu$m, except for the hottest stars where the flux at B was estimated 
using the effective wavelength approach, and linear interpolations used
from zero wavelength to B, and from B to the start of the red spectrum.
For the two stars without optical spectra the contribution in this region
was estimated from stars of similar temperature and metallicity.
For the hotter of these two stars, LHS 5327, there is a relatively large 
uncertainty in the total flux due to the lack of optical data; for this 
star the uncertainty in total flux is 10\%, leading to an uncertainty of 
0.10mag in the bolometric correction and 0.05dex in log$_{10}L/L_{\odot}$.
For the rest of our sample the uncertainties are 5\%, 0.05 mag and 0.02dex,
respectively.

Figure 8 plots K--band bolometric correction against I$-$K color.
We have included the results from \cite{l96}.  The approximate metallicities
of the stars are indicated, based on kinematic population. Model sequences are 
overlaid as dashed lines, where again these model calculations have not been 
upgraded to include the new TiO or H$_2$O linelists but still match the 
observations well.
The metal--poor stars are confined to small values of
BC$_K$ due to the onset of pressure--induced H$_2$ opacity, reducing the
flux at K.  For the disk stars the relationship between K--band bolometric 
correction and I$-$K color can be well represented by the cubic polynomial:
$$ BC_K =-2.741 +  5.452(I-K) - 1.824(I-K)^2 + 0.211(I-K)^3 $$
for 1.8 $\le$ I$-$K $\le$ 3.3.
This fit is indicated by the solid line in Figure 8.

\section{Models and Synthetic Spectra, and Comparison Process}

We have calculated the models presented in this paper using our multipurpose
model atmosphere code Phoenix, version 10.7. Details of the code and the
general input physics setup are discussed in the description of the
NextGen grid  of model atmospheres presented by \cite{all98b} and
references therein. The model atmospheres presented here were
calculated with the same general input physics as the NextGen models. However,
a change of the linelists has significant impact on the model structure and
synthetic spectra.  The most important difference from our NextGen grid
is the replacement of TiO and
H$_2$O linelists with the newer linelist calculated by the NASA--AMES group,
\cite{tio} for TiO (about 175 million lines of 5 isotopes) and
\cite{h2o} for H$_2$O (about 350 million lines in 2 isotopes).
Our combined molecular line list includes about 550 million molecular lines.
These lines are treated with a direct opacity sampling technique where each
line has its individual Voigt (for strong lines) or Gauss (weak lines) line
profile, see \cite{all98b} and references therein for details. The number of
lines selected by this procedure depends on the the model parameters.
In addition to the new line data, we have also included dust formation and
opacities in the models used in this paper. However, the lowest effective
temperatures of the stars we consider here are slightly above the regime were
dust formation and opacities are important.
A complete description of the models and the differences to the NextGen
models will be given in Allard \& Hauschildt (1999, in preparation).

The fitting of the synthetic to the observed spectra was done using an
automatic IDL program. First, the resolution of the synthetic spectra is
reduced to that of each individual observed spectrum and the spectra are
normalized to unit area for scaling.  The comparison was done using a model
atmosphere grid that covers the range 1500~K$\le$ T$_{eff} \le$ 4000~K, 
3.5 $\le$ log~g $\le$ 5.5 and $-1.5 \le$ [m/H] $\le 0.0$, with a total of 
221 model atmospheres.
For each observed spectrum we then calculate the $\chi^2$ value for the
comparison with all synthetic spectra in the grid. In order to avoid known
problematic spots in either the observations or the synthetic spectra, the
wavelength ranges 1.35---1.5$\mu$m and 1.8---1.95$\mu$m were excluded from the
comparison, however, tests showed this did not significantly change the
results. We selected the models that resulted in  the lowest 3---5 $\chi^2$ 
values as
the most probable parameter range for each individual star. The ``best'' value
was then chosen by visual inspection. This procedure allows a rough estimate of
the probable range in the stellar parameters. Note that systematic errors due 
to missing or incomplete opacity sources are not eliminated, however
investigations of the different linelists available for TiO and
H$_2$O indicate that differences in the treatment of these opacity sources 
effect the implied effective temperatures in opposite senses
(\cite{ahs99}) --- i.e. the systematic errors should be small.

\section{Results of Comparison of Data and Models}

The automatic comparison described in the previous section
resulted in $\chi ^2$ values between 0.02 and 0.10, with an average value
of 0.05.  The best fits were inspected by eye and in some cases a match
with a slightly higher $\chi ^2$ value than minimum was adopted.  We did
not try to match the bottom of the water bands, but did look at the depth
of the CO and TiO bands, and tried to match the overall ``continuum'' in
the optical and infrared regimes.  For the metal--poor stars the flatness
of the infrared continuum could be used to constrain the derived
metallicities. 

Based on visual inspection and the $\chi ^2$ values, the uncertainty in
the derived values of effective temperature
is $\pm 100$~K, and in metallicity ([m/H]) $\pm 0.25$~dex.  Gravity
could not be well constrained by data of this relatively low resolution
--- spectra generated with log~g values of 5.0 $\pm 0.5$~dex all
matched the data well.  Table 4 lists the derived parameters for the sample. 
For the two stars in common with \cite{l96} who used earlier NextGen
atmospheres, the agreement is within the quoted 150~K errors for that work
(for LHS 57 the temperature derived here is identical and for LHS 377 
it is 150~K cooler).

Figure 9 shows comparisons of synthetic and observed spectral energy 
distributions for approximately solar-metallicity stars with effective 
temperatures of 3600~K (LHS 65), 3100~K (LHS 421) and 2600~K (LHS 523). 
Also shown is a more metal--poor star with T$_{eff}=$3200~K (LHS 3061).
The fits are good except for the coolest temperatures where problems with
the water opacity became apparent, and where details of grain formation
and settling may become important. The strength of the FeH line at 1$\mu$m
is also overestimated in the models.   These new models have resolved the 
discrepancy between the infrared and optical regions seen by \cite{vit97}
for the eclipsing binary CM Draconis (LHS 421); Figure 9 shows that the
entire energy distribution is well matched by a model with T$_{eff}=$3100~K
and [m/H]$=-0.5$.

Figures 10---12 show derived T$_{eff}$ as a function of various colors.  
Symbols indicate the metallicity implied by the energy distribution 
comparison (given in Table 4) --- metal--poor stars are bluer at a constant 
temperature, except in V$-$I at T$_{eff} \le 2900~$K.
The solid lines are model--predicted temperature:color relationships; 
the apparent offset for the infrared colors of the disk stars 
is probably due to the
remaining problems with water opacity,  which is more significant at cooler
temperatures.  The error in color due to 
absolute calibration or observational error is smaller than the apparent offset.
We note that the observed trend of T$_{eff}$ with V$-$I and I$-$K colors
agrees with that implied by our earlier work using the NextGen atmospheres
(\cite{l96}, Figure 17) but the agreement between observation and theory
for V$-$I is much improved with the new TiO linelist.

Diameters have been derived for the stars in the sample in two ways.  
One was to use the scaling factors necessary to 
match the synthetic stellar surface spectra to that observed at the Earth,
which requires use of the trigonometric parallax and which results in errors
in diameter of around 10\% --- or that in the parallax if that is larger.  
The other method was to use Stefan's Law to derive diameter from
the observed stellar luminosity at the Earth and the  effective temperature.
Here the largest uncertainty comes from that in T$_{eff}$ which enters
as the fourth power; the typical uncertainty in diameter is then 13\%.
These two determinations give values for diameter that agree very well ---
to typically 1.5\%.  Diameters are also given in Table 4.

Figure 13 shows diameter (determined from scaling factor) versus effective 
temperature with symbols 
indicating the metallicity implied by the energy distribution 
comparison.    Metal--poor stars have a smaller diameter for a constant 
temperature.  Open symbols are known multiple systems which would have
larger diameters;  typical error bars as well as larger individual errors 
are indicated.  The dashed lines are structural model predictions from
\cite{bar97,cha97} (which use the NextGen  atmospheres  without the improved 
linelists) for an age of 10~Gyr.  The models for 1~Gyr are not significantly 
different.  The agreement is good except
for known multiple systems and for LHS 1183 (a young flare star), LHS 135, 
LHS 2945, LHS 5327 and LHS 549 (an old variable star),
which also seem to have too large a diameter.
As far as we are aware none of these are known to be multiple, and 
LHS 549 has been searched for close companions using speckle  by \cite{lei97}
with a negative result.  These five stars merit further study for 
multiplicity.

\section{Conclusions}

We have obtained 1---2.5$\mu$m spectra for 42 disk and halo M1--M6.5
dwarfs.  These data have been combined with new or published optical
spectra, and energy distributions derived by flux calibrating and
combining the individual spectral regions for each object.  Bolometric
luminosities
have been determined and a relationship between bolometric magnitude and
I$-$K color given.

The colors and energy distributions have been compared to synthetic
photometry and spectroscopy generated by an upgraded version of the
NextGen  models, the AMES-Dusty models (\cite{all98b}).  These models 
use more recent TiO and H$_2$O linelists, and include grain condensation and
extinction.  The agreement is much improved in the red region compared 
to earlier comparisons (e.g. \cite{l96}).
Problems remain with the match to the observed FeH features and also
to the water bands.  These problems are being addressed with
ongoing work to calculate more complete linelists.  Nevertheless we can
determine effective temperatures
for the sample to $\pm$100~K, metallicities to $\pm$0.25~dex, and
radii to typically 10\%.  Recent structural models by \cite{bar97,cha97}
agree with the luminosities and radii derived except for five stars
which may be previously unknown multiple systems:  LHS 1183, LHS 135, LHS 2945,
LHS 5327 and LHS 549.

\acknowledgments
We are very grateful to the staff at UKIRT, the University of Hawaii's
88--inch telescope, the Lowell Observatory and the Naval Observatory Flagstaff
Station for their assistance in obtaining the data presented in this paper.  
UKIRT, the United Kingdom 
Infrared Telescope, is operated by the Joint Astronomy Centre Hilo Hawaii
on behalf of the U.K. Particle Physics and Astronomy Research Council.  
FA acknowledges support from NASA LTSA NAG5-3435 and 
NASA EPSCoR grants to Wichita State University, and support
from CNRS.  PHH acknowledges partial support from the 
P\^ole Scientifique de
Mod\'elisation Num\'erique at ENS-Lyon.  This work was also supported in part 
by NSF grants AST-9417242, AST-9731450, and NASA grant NAG5-3505.  
Some of the calculations presented in
this paper were performed on the IBM SP and the SGI Origin 2000 of the UGA UCNS
and on the IBM SP of the San Diego Supercomputer Center (SDSC), with support 
from the National Science Foundation, on the Cray T3E of the NERSC with
support from the DoE, and on  the IBM SP2 of the French Centre National 
Universitaire Sud de Calcul (CNUSC) and the Cray T3E of the Commissariat a 
l'Energie Atomique (CEA). We thank all these institutions for a generous
allocation of computer time.

\clearpage 

\begin{figure}
\epsscale{.8}
\plotone{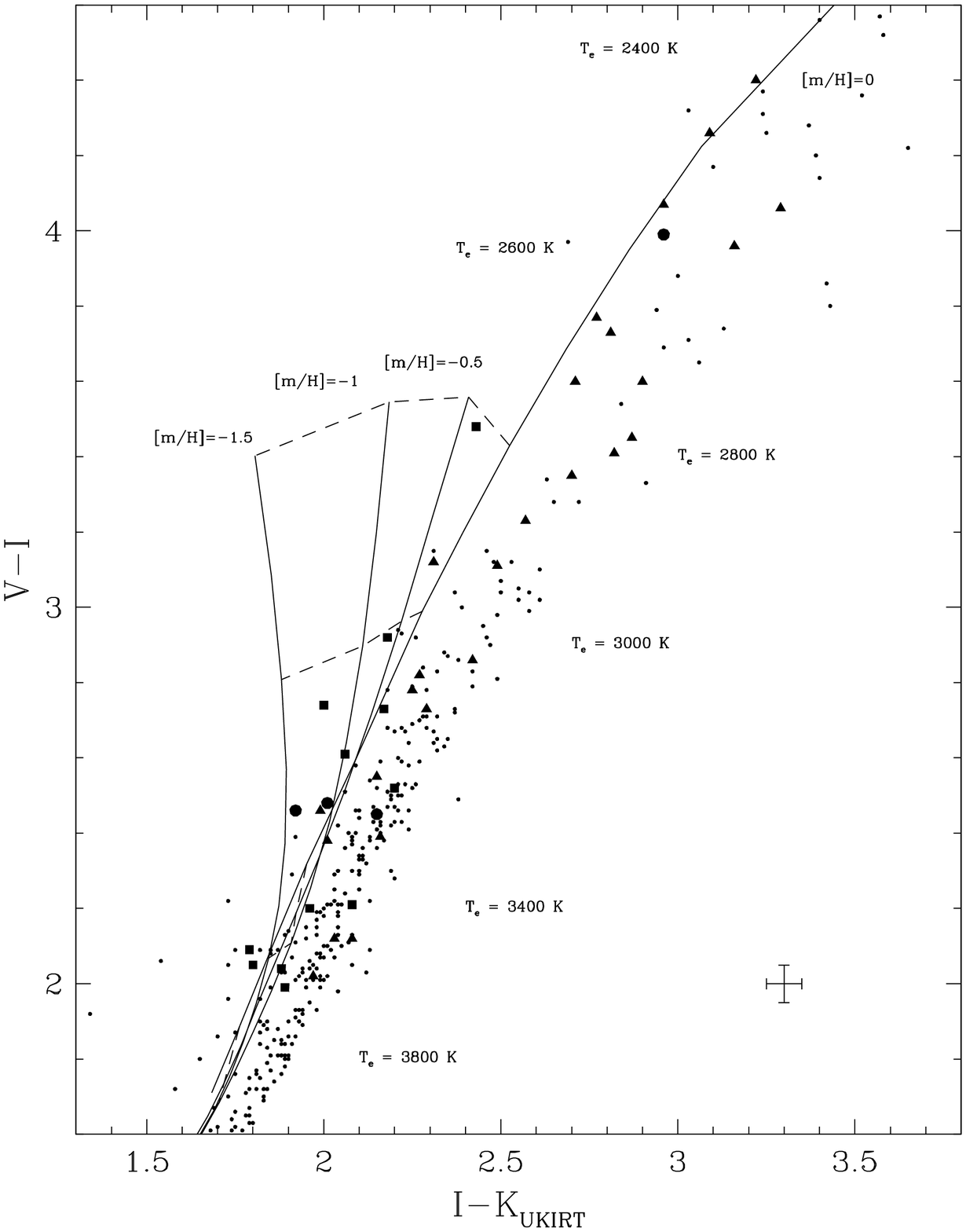}
\caption{V$-$I:I$-$K diagram with model sequences for metallicities 
 [m/H] $ = -1.5, -1, -0.5 $ and 0.  Dashed lines connect fixed
effective temperature values as indicated to the right.  Filled
symbols are
this work, where symbol shapes represent kinematic populations: squares
 --- halo, triangles --- disk, circles --- unknown.  Dots are stars from
\cite{l98} and from L92 (on the UKIRT JHK system). 
\label{fig1}}
\end{figure}

%\figcaption[fig1.ps]{V$-$I:I$-$K diagram with model sequences for
%metallicities [m/H] $ = -1.5, -1, -0.5 $ and 0.  Dashed lines connect
%fixed
%effective temperature values as indicated to the right. Filled symbols
%are
%this work, where symbol shapes represent kinematic populations: squares   
%--- halo, triangles --- disk, circles --- unknown.  Dots are stars from
%\cite{l98} and from L92 (on the UKIRT JHK system).
%\label{fig1}}

\begin{figure}
\epsscale{.8}
\plotone{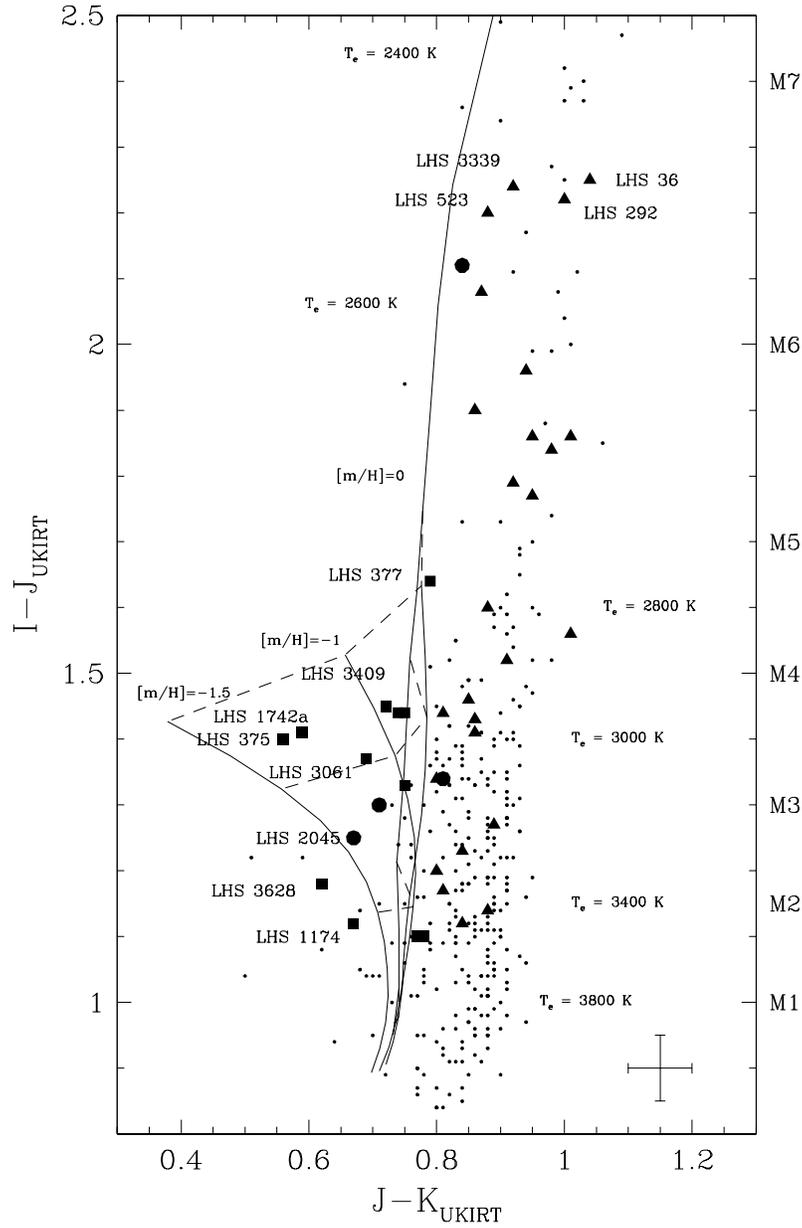}
\caption{I$-$J:J$-$K diagram with model sequences  using the same
symbols as in Figure 1. Spectral types based on I$-$J (L92) are shown.
\label{fig2}}
\end{figure}

%\figcaption[fig2.ps]{I$-$J:J$-$K diagram with model sequences  using the
%same  symbols
%as in Figure 1. Spectral types based on I$-$J (L92) are shown.
%\label{fig2}}

\begin{figure}
\epsscale{.8}
\plotone{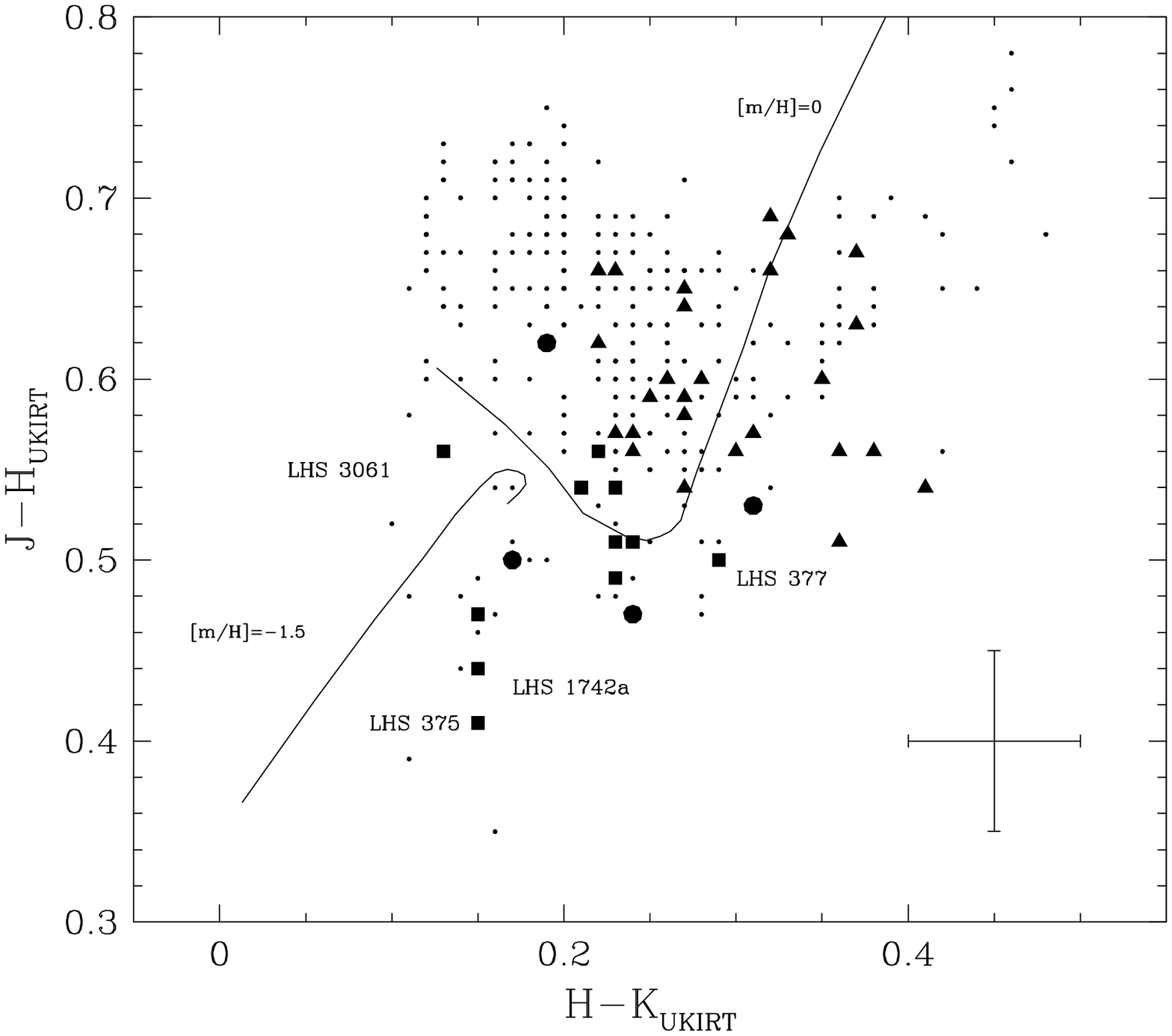}
\caption{J$-$H:H$-$K diagram with model sequences using the same  symbols 
as in Figure 1.
\label{fig3}}
\end{figure}

%\figcaption[fig3.ps]{J$-$H:H$-$K diagram with model sequences using the
%same  symbols as in Figure 1.
%\label{fig3}}

\begin{figure}
\epsscale{.8}
\plotone{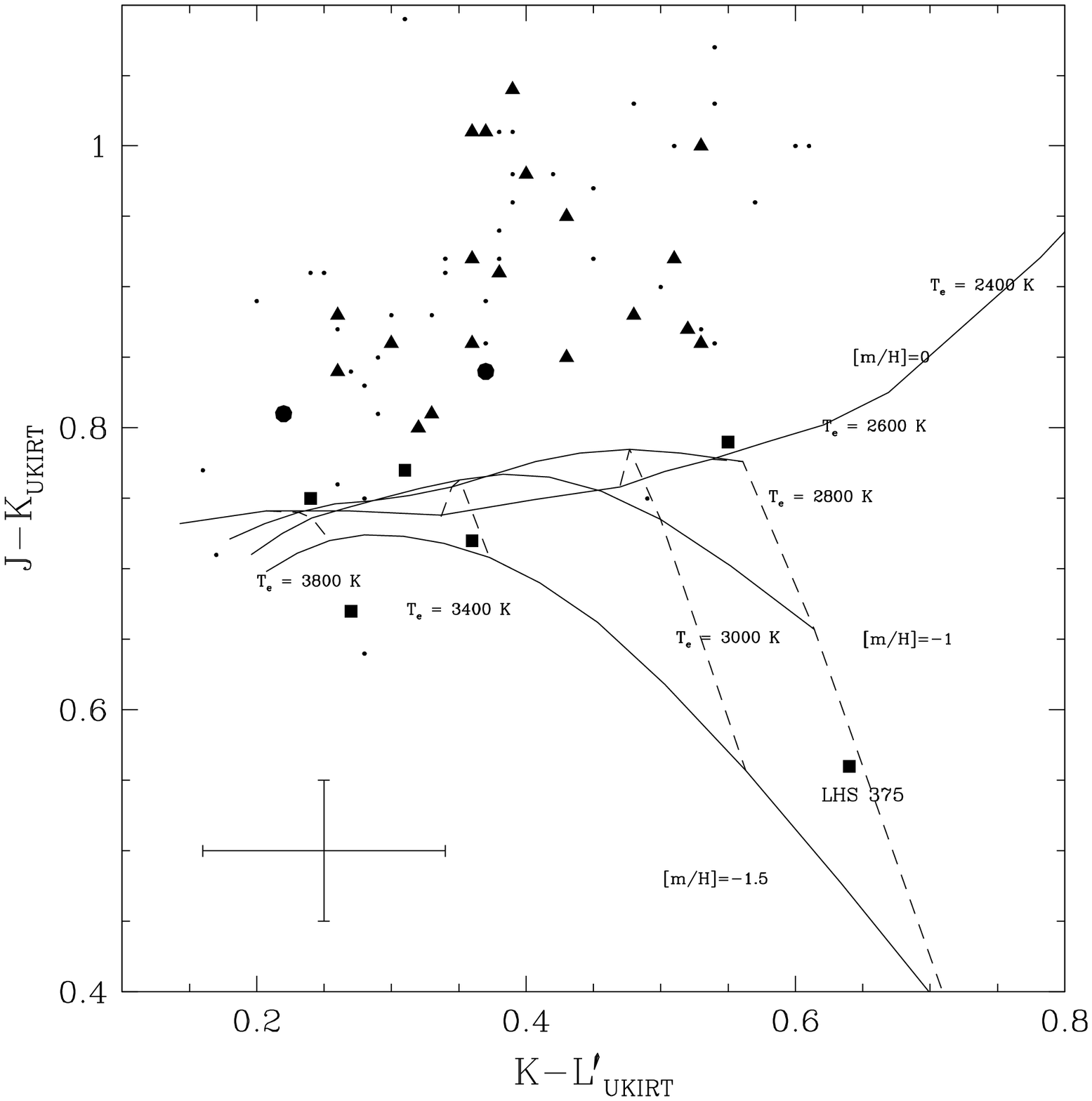}
\caption{J$-$K:K$-$L$^{\prime}$ diagram with model sequences using the
same  symbols as in Figure 1.
\label{fig4}}
\end{figure}

%\figcaption[fig4.ps]{J$-$K:K$-$L$^{\prime}$ diagram with model sequences
%using the same symbols as in Figure 1.
%\label{fig4}}

\begin{figure}
\epsscale{.8}
\plotone{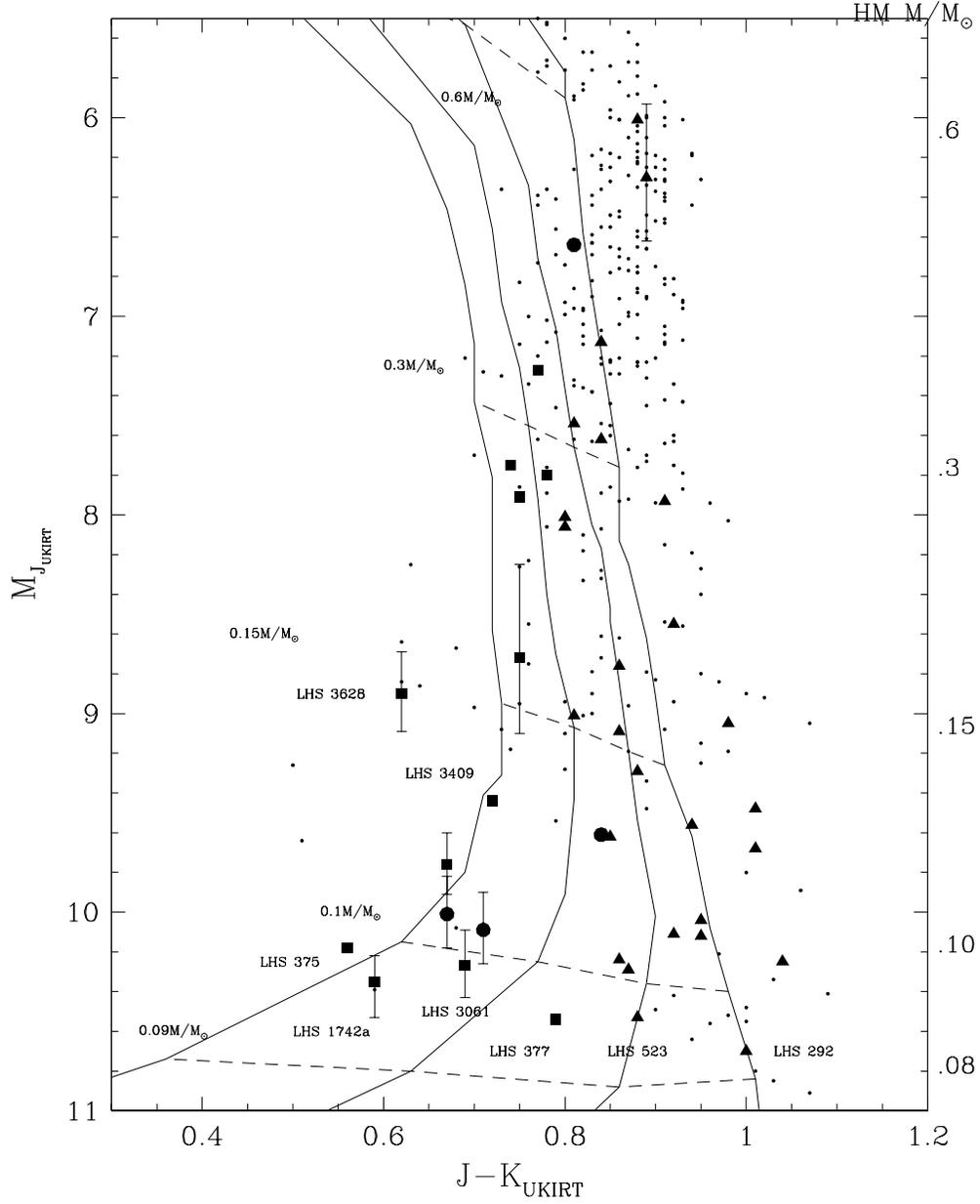}
\caption{M$_J$:J$-$K with symbols  as in Figure 1.
Empirical masses for the disk stars from Henry \& McCarthy 1993 are given
on the right axis.  Solid lines are isochrones from \cite{bar97,cha97}
for age 10~Gyr and metallicities from left to right: [m/H]
$= -1.5, -1.0, -0.5, 0.0$.   Dashed lines connect fixed mass values, 
as indicated.
\label{fig5}}
\end{figure}

%\figcaption[fig5.ps]{M$_J$:J$-$K with symbols  as in Figure 1.
%Empirical masses for the disk stars from Henry \& McCarthy 1993 are given
%on the right axis.  Solid lines are isochrones from \cite{bar97,cha97}   
%for age 10~Gyr and metallicities from left to right: [m/H]
%$= -1.5, -1.0, -0.5, 0.0$.  Dashed lines connect fixed mass values,
%as indicated.
%\label{fig5}}

\begin{figure}
\plotfiddle{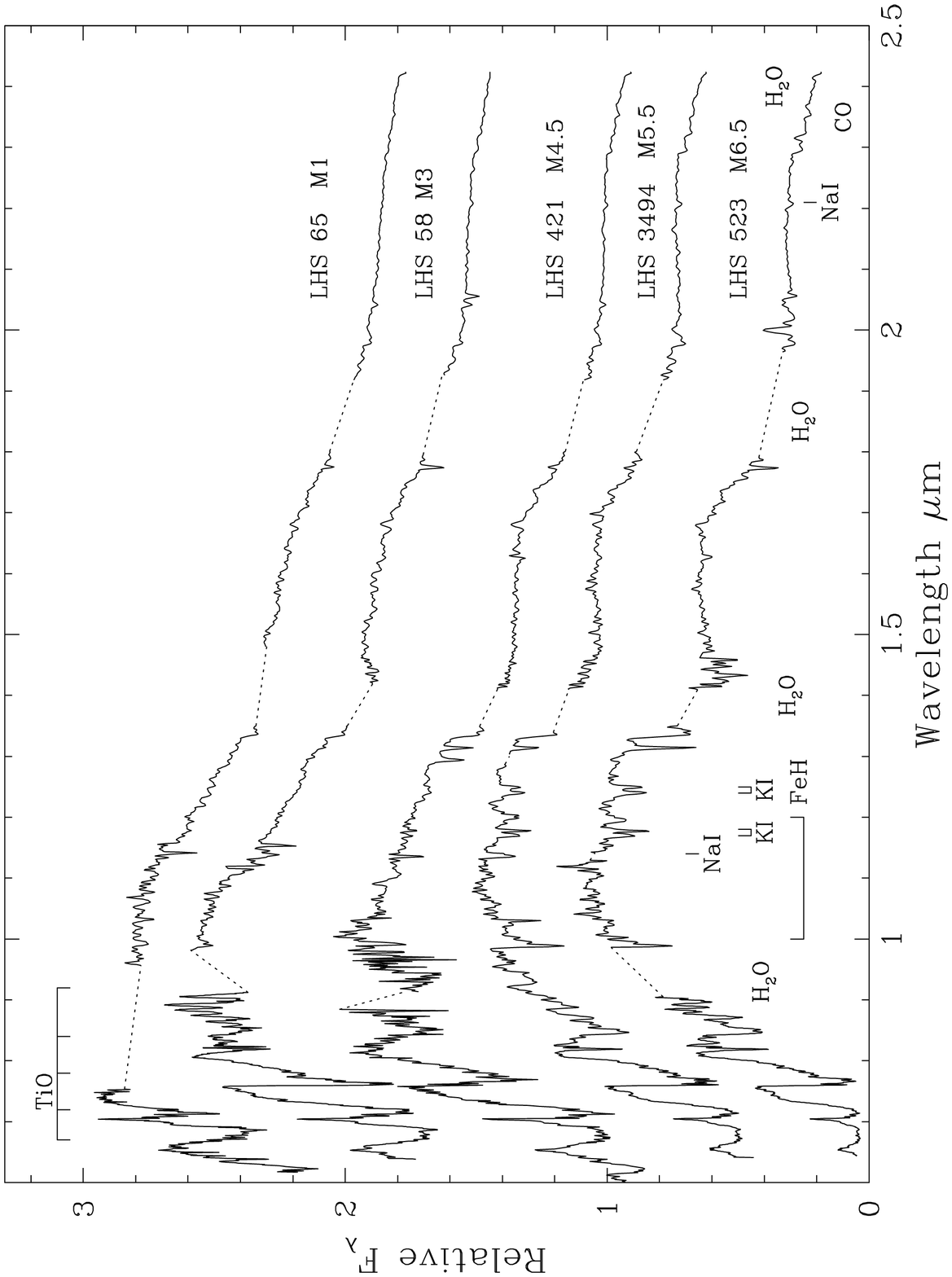}{10truecm}{-90}{75}{80}{-270}{450}
\caption{Spectral sequence for M dwarfs where the spectra have been
normalised to the flux at 1.2$\mu$m and offset.
\label{fig6}}
\end{figure}

%\figcaption[fig6.ps]{Spectral sequence for M dwarfs where the spectra
%have
%been normalised
%to the flux at 1.2$\mu$m and offset.
%\label{fig6}}

\begin{figure}
\plotfiddle{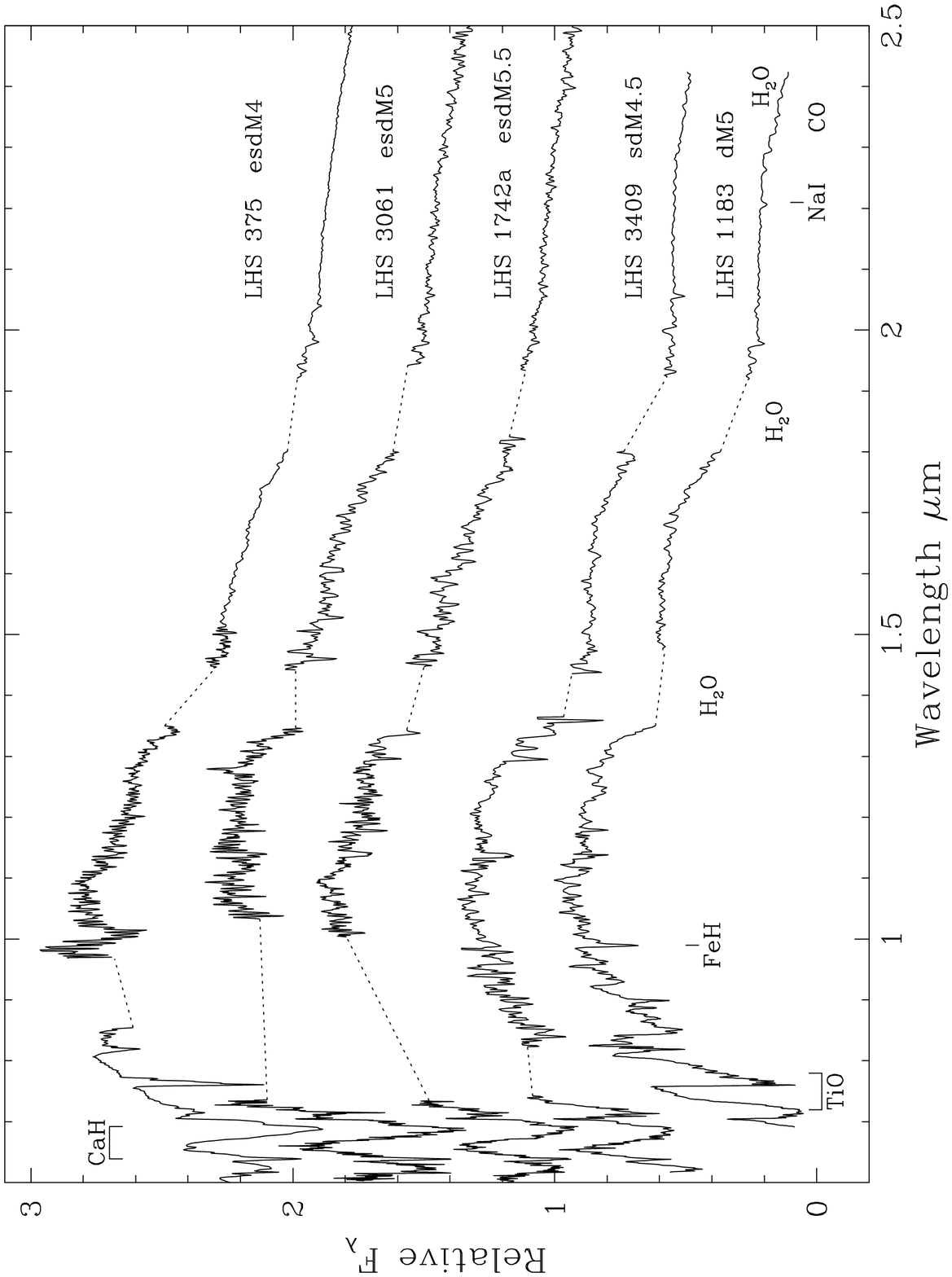}{10truecm}{-90}{75}{80}{-270}{450}
\caption{Spectral sequence for subdwarfs where the spectra have been
normalised to the flux at 1.2$\mu$m and offset.  
\label{fig7}}
\end{figure}

%\figcaption[fig7.ps]{Spectral sequence for subdwarfs where the spectra
%have been normalised
%to the flux at 1.2$\mu$m and offset.
%\label{fig7}}

\begin{figure}[t!]
%\plotone{fig8.ps}
\plotfiddle{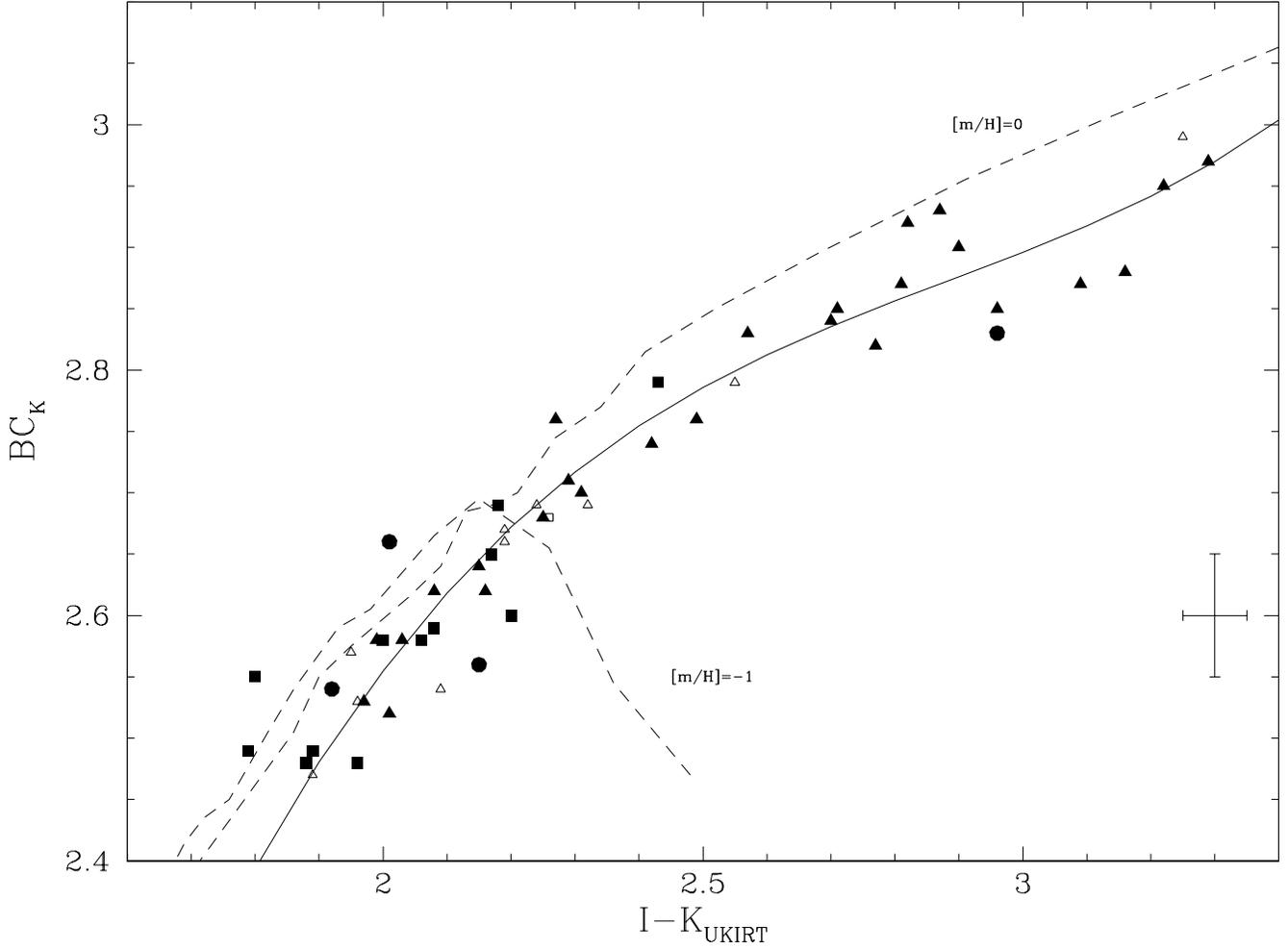}{10truecm}{-90}{70}{70}{-270}{400}
\caption{Bolometric correction at K versus I$-$K.  Filled symbols are
this work, open are from \cite{l96}.  Symbol shape represent kinematic 
populations: squares --- halo, triangles --- disk, circles --- unknown.
Model sequences for [m/H]$=$0 and   [m/H]$=-$1 are overlaid as dashed
lines.  The solid line is the empirical fit to the disk stars.
\label{fig8}}
\end{figure}

%\figcaption[fig8.ps]{Bolometric correction at K versus I$-$K.  Filled
%symbols are
%this work, open are from \cite{l96}.  Symbol shape represent kinematic
%populations: squares --- halo, triangles --- disk, circles --- unknown.
%Model sequences for [m/H]$=$0 and   [m/H]$=-$1 are overlaid as dashed
%lines.  The solid line is the empirical fit to the disk stars.
%\label{fig8}}

\begin{figure}
\plotfiddle{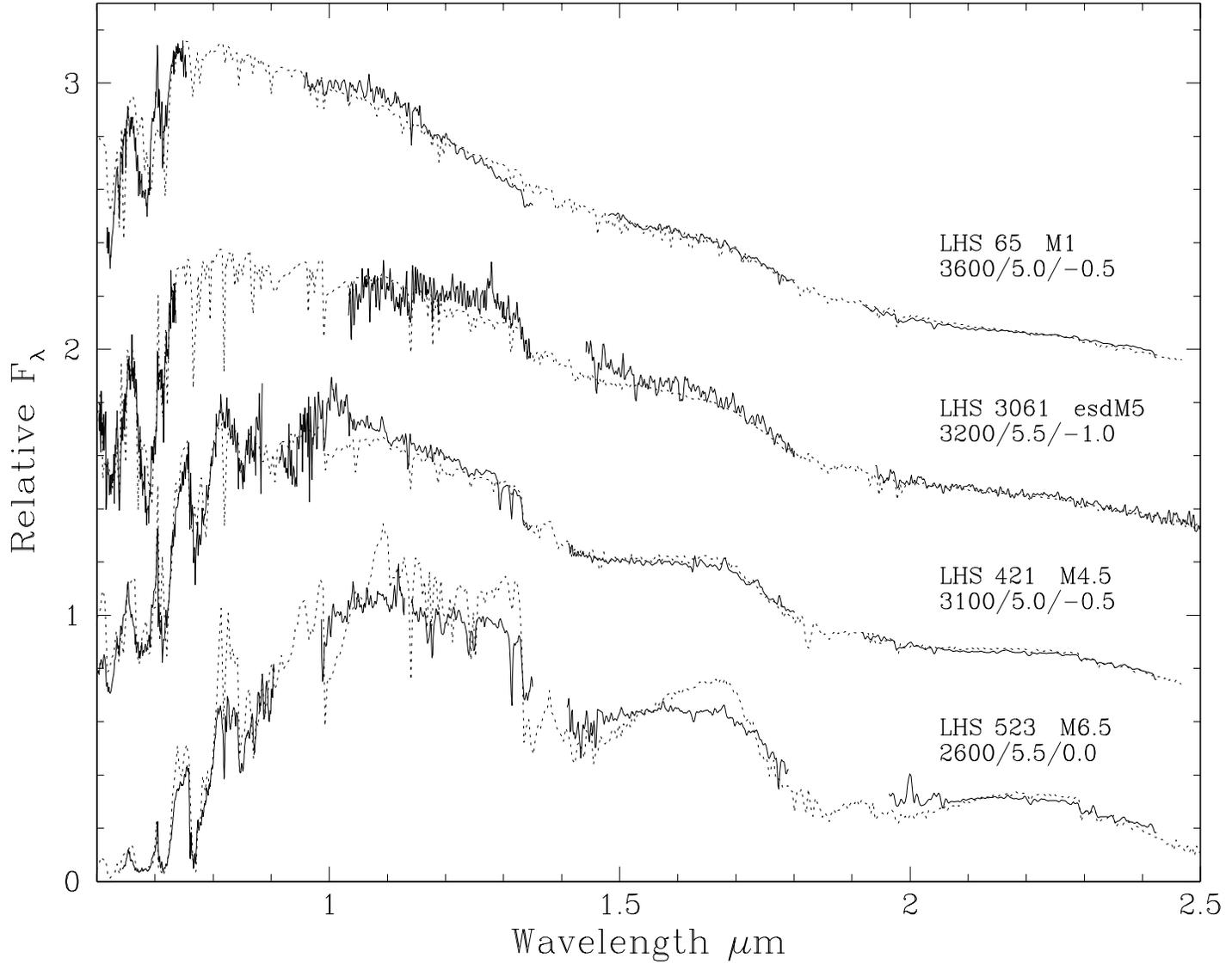}{10truecm}{-90}{75}{80}{-270}{450}
\caption{Comparison of synthetic (dashed line) and observed (solid line)
spectral energy distributions, 
where the spectra have been normalised 
to the flux at 1.2$\mu$m and offset.  The best fit model parameters
T$_{eff}$/log~g/[m/H] are given for each star.
\label{fig9}}
\end{figure}

%\figcaption[fig9.ps]{Comparison of synthetic (dashed line) and observed
%(solid line) spectral energy distributions,
%where the spectra have been normalised
%to the flux at 1.2$\mu$m and offset.  The best fit model parameters
%T$_{eff}$/log~g/[m/H] are given for each star.
%\label{fig9}}

\begin{figure}
\epsscale{.8}
\plotone{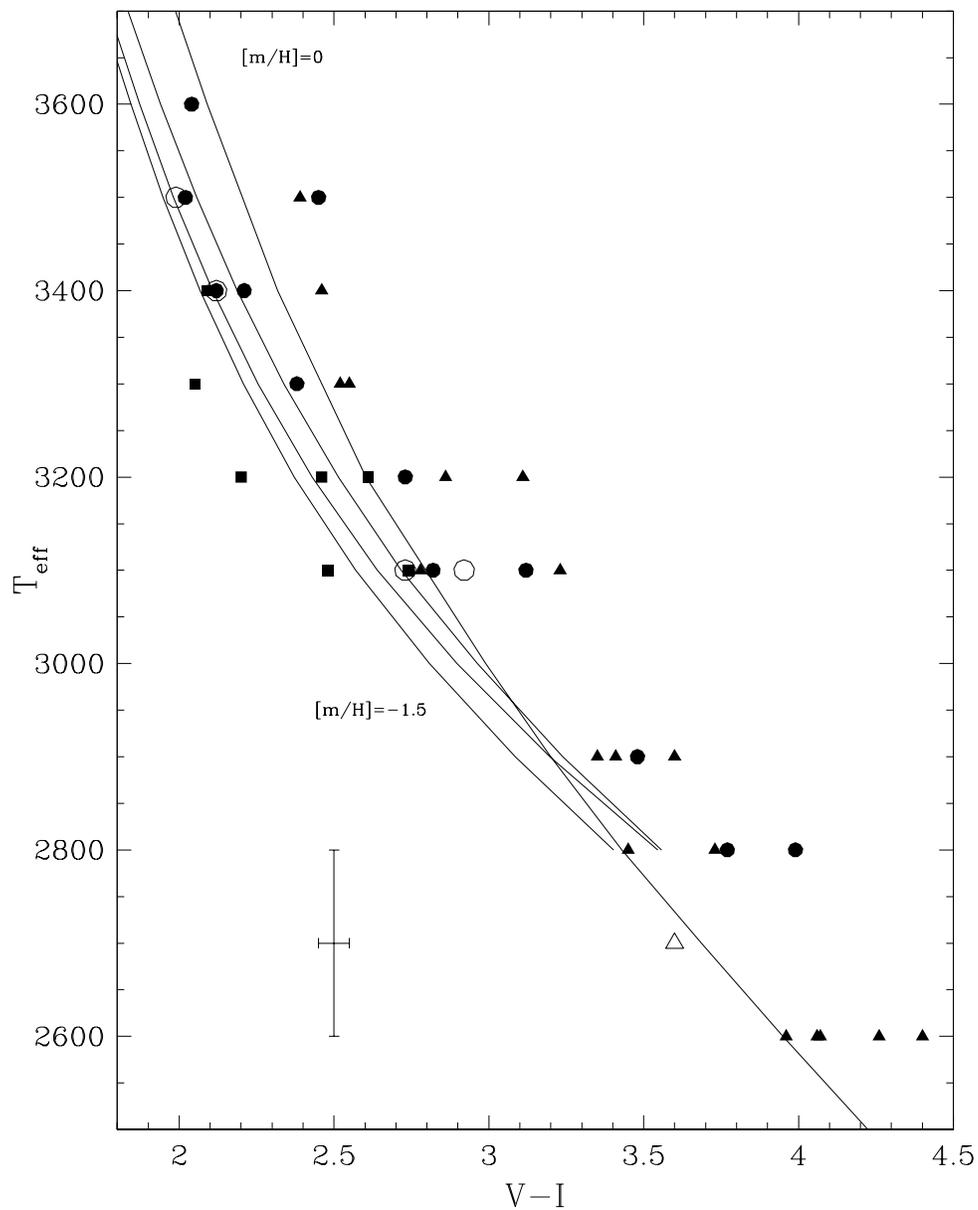}
\caption{ T$_{eff}$ as a function of V$-$I.  
Symbols indicate the metallicity implied by the energy distribution:
squares --- [m/H]$=-$1.0, circles --- [m/H]$=-$0.5, triangles ---
[m/H]$=$0.  Open symbols are multiple systems.  Model sequences for
[m/H]$=$0, $=-$0.5, $=-$1.0 and  $=-$1.5 are also shown where for hotter
stars decreasing metallicity at a constant temperature results in bluer 
V$-$I. 
\label{fig10}}
\end{figure}

%\figcaption[fig10.ps]{T$_{eff}$ as a function of V$-$I.
%Symbols indicate the metallicity implied by the energy distribution:
%squares --- [m/H]$=-$1.0, circles --- [m/H]$=-$0.5, triangles ---
%[m/H]$=$0.  Open symbols are multiple systems.  Model sequences for
%[m/H]$=$0, $=-$0.5, $=-$1.0 and  $=-$1.5 are also shown where for hotter
%stars decreasing metallicity at a constant
%temperature results in bluer V$-$I.
%\label{fig10}}

\begin{figure}
\epsscale{.8}
\plotone{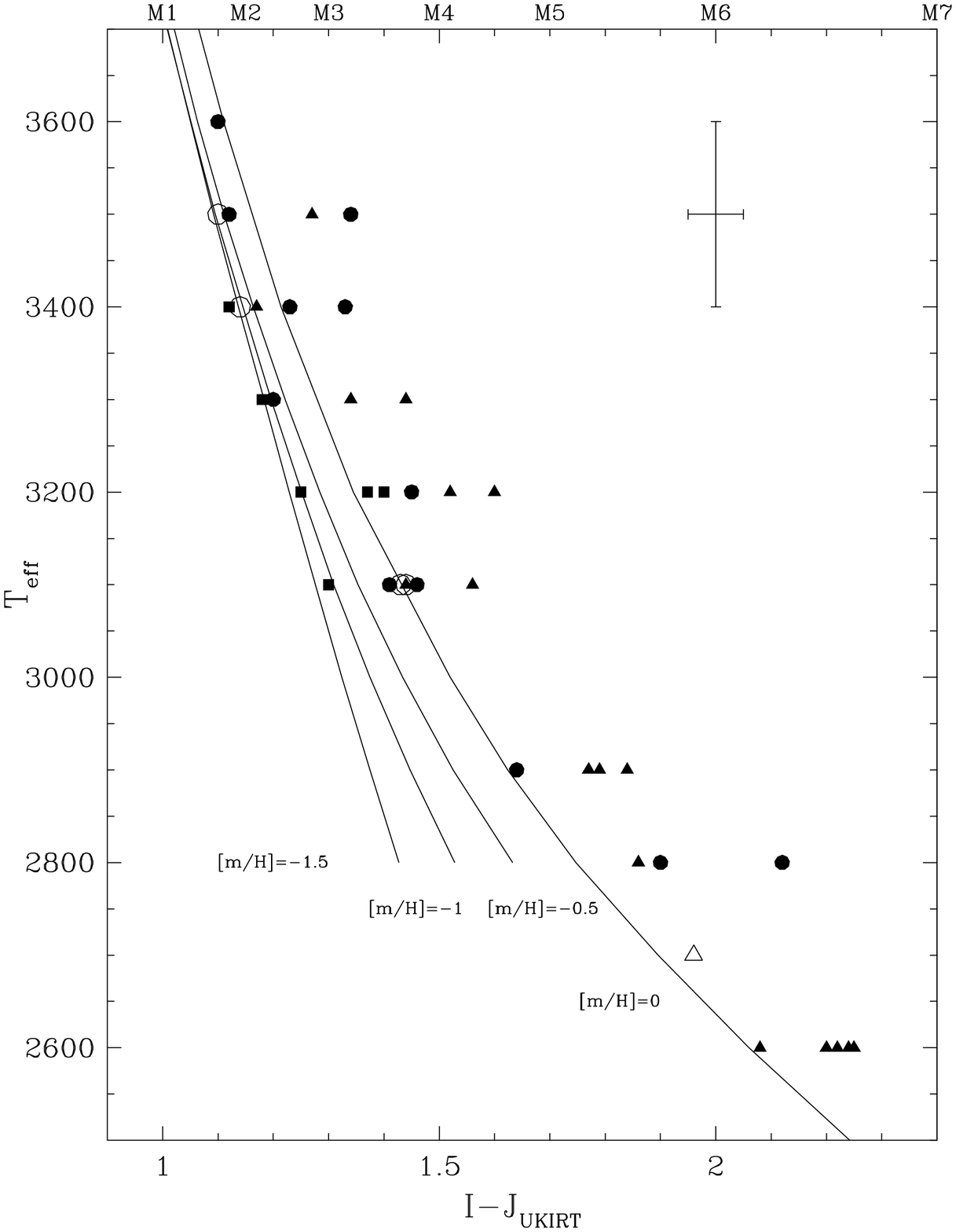}
\caption{ T$_{eff}$ as a function of I$-$J.  
Symbols are as in Figure 10.
\label{fig11}}
\end{figure}

%\figcaption[fig11.ps]{T$_{eff}$ as a function of I$-$J.
%Symbols are as in Figure 10.
%\label{fig11}}

\begin{figure}
\epsscale{.8}
\plotone{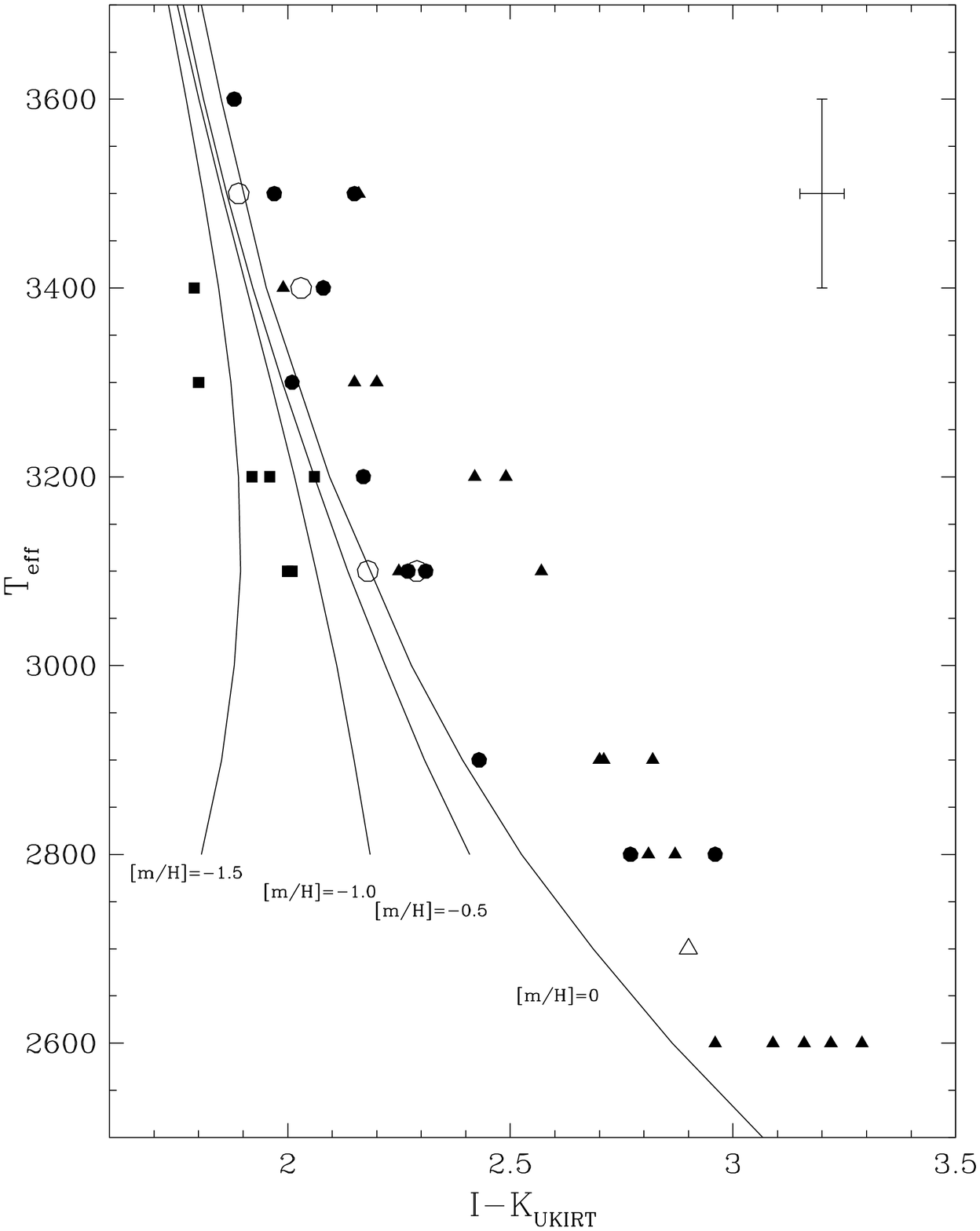}
\caption{ T$_{eff}$ as a function of I$-$J.  
Symbols are as in Figure 10.
\label{fig12}}
\end{figure}

%\figcaption[fig12.ps]{T$_{eff}$ as a function of I$-$J.
%Symbols are as in Figure 10.
%\label{fig12}}

\begin{figure}
\plotfiddle{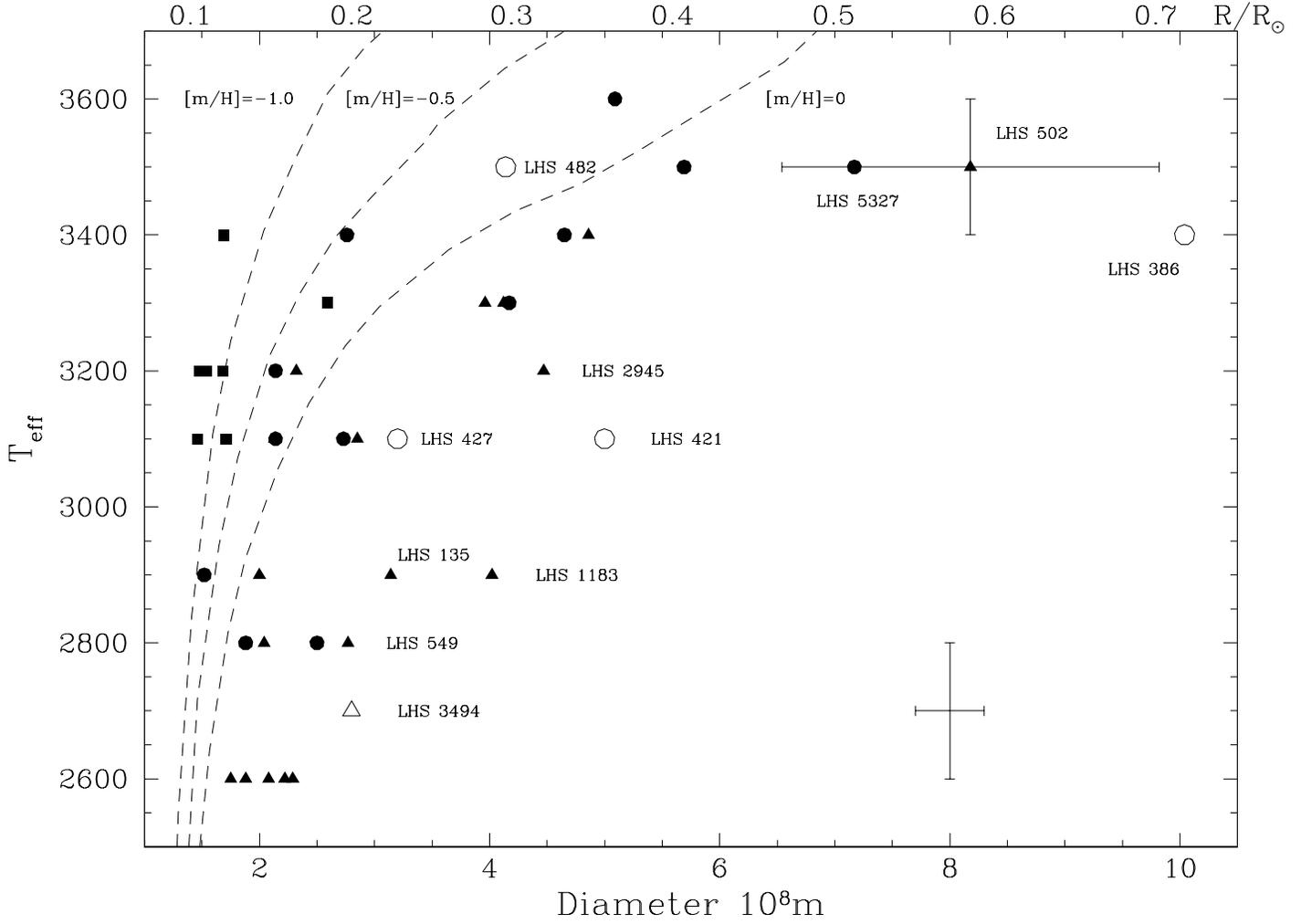}{10truecm}{-90}{70}{70}
{-270}{400}
\caption{Diameter (derived by scaling) as a function of  T$_{eff}$.
Symbols are as in Figure 10 and dashed lines are structural model
calculations from \cite{bar97,cha97}.  Stars
with apparently large diameters are identified, as is a star with a large
uncertainty due to parallax error.
\label{fig13}}
\end{figure}

%\figcaption[fig13.ps]{Diameter (derived by scaling) as a function of
%T$_{eff}$.  Symbols are as in Figure 10 and dashed lines are structural
%model calculations from \cite{bar97,cha97}.  Stars with apparently large
%diameters are identified, as is a star with a large uncertainty due to
%parallax error.
%\label{fig13}}

\end{document}